Scott Humr[a]* and Mustafa Canan[a]

[a]*Information Sciences, Naval Postgraduate School, Monterey, CA, USA*

*corresponding author: scott.humr@nps.edu*

# You Can't Get There From Here:
# Redefining Information Science to address our socio-technical futures

**Abstract**

Current definitions of Information Science are inadequate to comprehensively describe the nature of its field of study and for addressing the problems that are arising from intelligent technologies. The ubiquitous rise of artificial intelligence applications and their impact on society demands the field of Information Science acknowledge the socio-technical nature of these technologies. Previous definitions of Information Science over the last six decades have inadequately addressed the environmental, human, and social aspects of these technologies. This perspective piece advocates for an expanded definition of Information Science that fully includes the socio-technical impacts information has on the conduct of research in this field. Proposing an expanded definition of Information Science that includes the socio-technical aspects of this field should stimulate both conversation and widen the interdisciplinary lens necessary to address how intelligent technologies may be incorporated into society and our lives more fairly.

## Introduction

The need for the Information Sciences (IS) has never been more important for addressing the impacts of Artificial Intelligence (AI) on society. Besides humans, other cognitive-like systems such as AI, can act on information with consequential outcomes. From devastating impacts to our natural environments (Crawford, 2021), to algorithmic oppression (Noble, 2018) the need for IS practitioners that understand the full range of socio-technical aspects is crucial for promoting the best long-term outcomes of AI for society. For these reasons, current definitions that describe the content of IS are inadequate for characterizing what the field now encompasses. Though, scope has always been a problem in defining IS with some being too inclusive and others, not inclusive enough (Aspray, 2011). As AI becomes more capable and interacts with information on behalf of users, broadening the boundaries of IS are necessary to address the myriad of way this technology is impacting individuals and society.

Technologies such as AI and machine learning, while promising to make our lives easier, are also exploiting our predictable behaviors and marginalizing many groups (Birhane et al., 2021; O'Neil, 2017). As AI and similar technologies continue to permeate ever deeper into more aspects of human life, there is still little understanding of their long-term effects (Glikson & Woolley, 2020; Nishant et al., 2020; Siddarth et al., 2021). The desire to introduce figures of merit for every aspect of human life from calorie intake, steps taken, average resting heart rate to Amazon recommendations and continuous location monitoring, highlights the nature of emergent socio-technical aspects of life. The penchant for quantification and instrumentation of society alludes to the limitations of current definitions of IS that have been put forth over the last six decades. Moreover, AI is not only a substitutional technology but is infrastructural because it is transforming ways of life (Barley, 2020). Addressing these shortcomings requires a more comprehensive definition of IS to expand the basis of analysis in this field to place renewed emphasis on consideration of the additional enviro-social aspects of AI and similar intelligent technologies.

We address these shortcomings by first outlining AI with its environmental and social impacts, both at the individual and societal levels. Then, we show that IS provides an ideal socio-technical framing for understanding the nexus of humans and these technologies. Next, we discuss current definitions of IS and what they neglect. Finally, we discuss an amended definition of IS for the 21st century to foreground the impacts information has on both humans and societies. We close with implications for the field of IS.

## Impacts of AI

There is much controversy over the broad impacts of AI. From inscrutable algorithms and built-in biases to threats of unemployment, AI is portending to upend domains that have been the sole purview of humans. Yet, understanding the implications of AI will take an interdisciplinary approach to fully appreciate the potential long-term effects of this technology. AI has become abstracted to the point of appearing as magic (Elish & Boyd, 2018). However, behind the edifice of AI lies a vast ecosystem reliant on massive amounts of natural resources (Jones & Easterday, 2022), energy (Victor, 2019), and inexpensive human labor (Perrigo, 2023) that often goes unrecognized. We briefly highlight these harms across these sectors.

### Environmental Impacts of AI

AI is more than just the smart technologies people have access to in the palm of their hands through such services as Siri, Alexa, ChatGPT, or Google Maps. AI is composed of a vast ecosystem that is sustained through the continuous extraction of natural resources such as coal, rare earth minerals, and other toxic chemicals (Crawford, 2021). Materials needed to sustain the ongoing hardware needs of storage, silicon chips, and energy are necessary to maintain the ever-increasing needs of a technology to support additional data and compute requirements for training more complex models (Magubane, 2023). Yet, such impacts are often unrecognized.

The impacts of AI on the environment are also measurable by their operations. Large AI models such as frontier models powering large language models (e.g., ChatGPT, Claude) that have captured the public's attention are run on equipment held within enormous data centers (Pascual, 2023). Recent reports suggest that such AI applications even have an outsized carbon footprint (Coleman, 2023; Saul & Bass, 2023; Sundberg, 2023). AI techniques such as deep learning (e.g., neural networks), may consist of billions of hyperparameters that require increased energy consumption to train these models (Tamburrini, 2022). For instance, some researchers have suggested that AI computational requirements are projected to outpace all power produced worldwide by 2040 (Hopkin, 2023). While such prognostication may seem hyperbolic, these impacts are garnering more attention and highlighting the harms AI imposes on society (Labbe, 2021; Lai et al., 2022). These and other harms must be taken into consideration by the IS community to address the unsustainable practices of rampant information technology proliferation and AI energy consumption at the expense of the environment.

### Social Impacts of AI

AI has many human impacts that affect not only the environment, but the social fabric itself as these technologies become more embedded in society (Selenko et al., 2022). While warnings of AI's environmental impacts are more readily measurable, the social harms caused by AI are becoming more evident (Acemoglu, 2021; Smuha, 2021). The harms of AI are happening at several levels.

### AI's Societal Harms

Unlike the environmental harms, the social harms caused by AI are equally concerning. Researchers have empirically found AI's influence on society from the dissemination of disinformation (Whyte, 2020), class discrimination based on race and gender (Buolamwini & Gebru, 2018), disabilities (Whittaker et al., 2019), to general fairness (Madaio et al., 2022). The challenge with societal impacts of AI points to how such harms flow to the individual level (Acemoglu, 2021). For instance, disinformation campaigns enabled by AI portend an exacerbation of current trends towards undermining faith in democratic processes (Brkan, 2019; Kaplan, 2020; Landon-Murray et al., 2019). This was seen most clearly in the 2016 U.S. Presidential Campaign with Cambridge Analytica's involvement using Facebook data to harness the power of AI for influencing targeted populations (Isaak & Hanna, 2018; Schippers, 2020). However, such harms are enabled through the collection of massive quantities of data.

People provide the fuel for AI through the abundant amount of data they generate interacting with smart technologies today. Organizations such as Google, Facebook, and Amazon harvest personal data by offering free software in exchange for internet browsing behaviors, purchase proclivities, and location data (Freedman, 2023; Li et al., 2019). Companies and governments can then use this data to make decisions, but potentially perpetuate bias against large classes of people. A most often cited example of this is the Correctional Offender Management Profiling for Alternative Sanctions (COMPAS). COMPAS uses data to calculate recidivism risk, but empirical research demonstrated that the software program used by many U.S. states had a high number of false positives towards African American offenders (Mehrabi et al., 2021). These and other well-documented behaviors of AI systems cast a long shadow of harms toward society that IS should be positioned to address.

### AI's Individual Harms

Societal harms such as environmental degradation and marginalization of certain groups of people are ultimately felt at the individual level. Yet, such harms go even deeper to uphold the Potemkin edifice of AI wonders. Several studies and exposés have revealed the uncanny success behind much of AI's recent display of performance. Such research has revealed the outsourcing of AI success through exploitation of cheap labor from around the world (Altenried, 2019). What occidental peoples experience as the utopia-like capabilities of AI anticipating their every wish; in stark contrast, exploited populations live in a dystopia working long hours moderating obscene content and tagging abusive and sexual explicit images to uphold the appearance of something truly intelligent (Roberts, 2021). This dystopian ecosystem is upheld through what Lazzarato calls immaterial labor where risks are socialized, and profits are privatized (Steinhoff, 2019; Three Minute Theory, 2020). Such exploited workers often suffer physical harms such as anxiety, depression, and post-traumatic stress from viewing such content (Williams, 2022). Yet, many such workers are compelled or even forced to stay in these harmful jobs because of the lack of other opportunities (Jones, 2021).

Such harms do not stop behind the scenes of AI curation practices. Previously mentioned AI-enabled programs like COMPAS directly impact the lives of individuals. Research has demonstrated several automated decision-making technologies that carry with it legal harms (Bayamlıoğlu & Leenes, 2018; O'Neil, 2017). In other cases, AI is shown to restrict a worker's ability to understand the underlying algorithms and limits an employee from operating effectively (Strich et al., 2021). Processing enormous amounts of data for AI can also lead to violations of personal privacy, unfair competition, and behavioral manipulation (Acemoglu, 2021; Hacker, 2020). For these reasons, IS needs to adopt a socio-technical foundation that can operate across different, competing paradigms to better understand how humans and societies are impacted by such technologies.

**IS *is* Socio-technical**

IS is a science that has not only lost its identity, it has also stagnated for some time. Arguably, IS has been in a slow decline for some years as sub-fields within the field break off into their own disciplines and university departments reorganize (Meadows, 2008). With the rise of personal computers and more recently smartphones, the field of IS became less about librarians organizing knowledge and information to everyone curating their own information (Gilchrist, 2014). Combined with ubiquitous search and now augmented with AI, IS may be suffering from a lack of a comprehensive identity (Benbasat & Zmud, 2003; Floridi, 2002). However, we argue that the discipline of IS may recover from its current shortcomings with a switch of focus to what should now be its core conceptual framework: socio-technical systems.

For most of the modern world, existence in the 21st century is inescapably socio-technical. A socio-technical system is defined as the interaction between people and technology. Saracevic (2009) characterizes IS as a dual focus with one eye toward the technical and system-oriented and an eye towards the individual and social aspects. Still, the far-reaching implications of AI on society entail both impacts to the environment and the humans who occupy these systems. As seen, AI crosses multiple levels of analysis from macrosocial (society) down to individuals. Similarly, socio-technical analysis covers three levels consisting of primary work systems, organizations, and the macrosocial level (Trist, 1981). Socio-technical systems theory is based on that the concepts of (1) organizations are a socio-plus technological whole; and (2) that the whole relates to its intended environment (Cummings & Markus, 1979). To cover the conceptual space necessary for improving IS research, IS researchers and practitioners need to foreground the socio-technical approach. Additionally, IS researchers should embrace the concept that IS is particularly a science of the artificial after Herbert Simon (1996) as Buckland (2012) aptly points out. The concept of the sciences of the artificial yields a human design process that bridges both the environmental, technical, and social by showing how different fields may harmonize under the big tent of IS. Figure 1 depicts how IS can bridge that gap of these two spheres. Still, to move IS beyond its stagnant state, it must first ground itself in a more comprehensive definition for the field.

**Figure 1**

IS and Sciences of the artificial

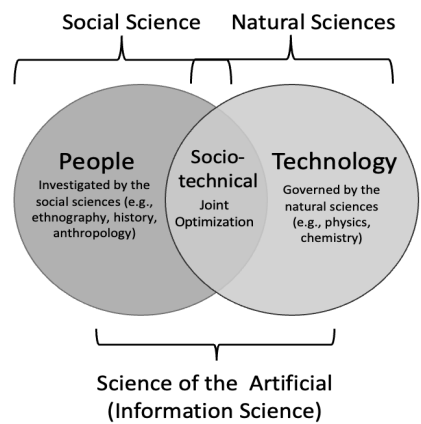

**IS Definitions**

Various definitions of IS were put forth since the advent of computers. The practice of IS goes back to at least the 1950s (Furner, 2015; Wersig & Neveling, 1975). Originally, IS was closely aligned, if not synonymous with library sciences (Meadows, 2008). Wellisch (1972) stated as far back as 1972 that at least 39 definitions of IS existed and were compared for common concepts where he concluded that no consensus existed. An often-repeated definition of IS was given by Borko (1968). Borko's definition is provided in Table 1 along with six more modern definitions for contrast and comparison. Tellingly, each definition lacks an explicit social component, save for Saracevic's (1999) definition.

Current definitions of IS are not only antiquated but lack the intersubjective aspects of today's technologies and the depth necessary for investigating such emergent phenomena. The inherent emphasis in Borko's definition on the technical-mechanical aspects of information has impeded progress and relevance to the field of IS (Buckland, 2012). These prominent definitions also neglect the rich understanding that emerges from the social and the human-cognitive interactions with AI technologies. However, the IS community acknowledged that while there was no agreed upon definition of IS at the time of his writing, researchers such as Roberts (1976) agreed upon the significance of the social aspects of information concepts and related phenomena in IS. Now likely lost to history, we may never know why the social aspects were never fully adopted into definitions of IS proper.

To rectify the shortcomings of these definitions and expand it to better encompass the socio-technical aspects, we argue that expanding the boundaries of the definition of IS will better align research efforts to truly become an interdisciplinary science for addressing intelligent technologies such as AI and its host of impacts to the environment and society.

**Table 1**

Sampling of IS Definitions

| Text Title | Definition | Source |
|---|---|---|
| Information science: What is it? | It is an interdisciplinary science that investigates the properties and behavior of information, the forces that govern the flow and use of information, and the techniques, both manual and mechanical, of processing information for optimal storage, retrieval, and dissemination. | (Borko, 1968) |
| An Introduction to Information Science | A more general approach would be to take information science as the study of the generation, preservation, and communication of information. | (Flynn, 1987) |
| Relationship with the natural sciences and knowledge engineering | Information science centers on the development of principles, laws, models, and theories that predict or explain information phenomena associated with natural artificial systems. Such systems include e.g., cells, molecules, organs, organisms, computers, organizations, communities, and atmospheric systems. | (Harmon, 1990) |
| Information Science | [I]nformation science is a field of professional practice and scientific inquiry addressing the problem of effective communication of knowledge records—"literature"—among humans in the context of social, organizational, and individual need for and use of information. | (Saracevic, 1999) |

| Paradigms in information science Steps towards a systemic paradigm for information science | Information science is regarded as the science dealing with structure and properties pertaining to information and communication, in addition to theories and methods for transfer, storage, recovery, evaluation and distribution of information. | (Adriaenssen & Johannessen, 2016) |
|---|---|---|
| Encyclopedia Britannica (online) | Information science, discipline that deals with the processes of storing and transferring information. It brings together concepts and methods from disciplines such as library science, computer science and engineering, linguistics, and psychology in order to develop techniques and devices to aid in the handling—that is, in the collection, organization, storage, retrieval, interpretation, and use—of information. | (Encyclopedia Britannica, 2021) |
| Introduction to Information Science | Information science is a field of study, with recorded information and documentation as its concern, focusing on the components of the information communication chain, studied through the perspective of domain analysis. | (Bawden & Robinson, 2022) |

**Recalibrating IS**

To argue for a new definition of IS requires an understanding of what current definitions lack. Endemic to much of the research in the early area of IS was an emphasis on positivism as a dominant epistemology for investigating phenomena (Walsham, 1995). Positivism assumes that reality is mind-independent and discoverable by a researcher (Geertz, 1973; Guba & Lincoln, 1994; Orlikowski & Baroudi, 1991). Arguably, the positivistic stance has led to an impoverished and partial understanding of information systems phenomena which is not sufficient (Orlikowski & Baroudi, 1991). Fortunately, the late 1980s saw a shift towards more research taking an interpretive perspective with the editors of MISQ (Management Information Systems Quarterly) requesting research from a more qualitative and integrationist perspective (Walsham, 1995). Researchers such as Orlikowski and Baroudi (1991) argued for a "plurality of research perspectives that could provide better understanding of phenomena in this field (p. 1). Therefore, to understand the rich nuances of socio-technical phenomena, researchers need to include an interpretive approach.

One aim of interpretive research is to give voice to the participants within a phenomenon (Orlikowski & Baroudi, 1991). The interpretive researcher starts with the assumption that access to reality is only through social constructions (Myers & Avison, 2002). Interpretive researchers work with the subjective meanings already in the world but are also faithful to avoid distorting these meanings (Schutz, 1972). These commitments allow the interpretive researcher to elicit complexity, ambiguity, and instability within their research (Orlikowski & Baroudi, 1991). Rich descriptions of phenomena allow researchers and others to build more accurate models of organizational and human behaviors (Daft & Wiginton, 1979). As a result, additional insights provided by interpretive research add a necessary interactive socio-component to the definition of IS.

Understanding the "properties and behavior of information, the forces that govern the flow and use of information" (Borko, 1968, p. 3) requires insights beyond a superficial level. Inquiry into some phenomena may require special equipment or measurement devices to record attributes, indicators, or variables the researcher deems significant. Unfortunately, many of the phenomena of interest in the socio-cognitive world of humans defy simple measurements. Recent research suggests that taking measurements or asking questions of subjects *creates* rather than records their behavior [emphasis added] (Busemeyer & Bruza, 2014). Constructs such as trust, level of commitment, and perseverance are difficult to quantify because truly objective measures and single referents do not exist. Yet, some of the most important aspects of culture and human interactions within technology are captured in such latent variables.

Lastly, other interpretive frameworks are necessary for exploring IS phenomena that capture underlying meanings not yet investigated such as critical research. For instance, critical research "focuses on the oppositions, conflicts, and contradictions in contemporary society, and seeks to be (Myers & Avison, 2002). Critical research also confronts "structures of oppression" (Kincheloe et al., 2017, p. 418). Moreover, qualitative approaches contain an exploratory component for uncovering an emergent story that would not have been predicted a priori (Kreiner et al., 2017). Therefore, qualitative inquiry, such as critical research, provides a deeper appreciation and understanding of phenomena such as AI that defy simplistic approaches while also augmenting conventional IS approaches.

## The IS Bridge: Sciences of the Artificial

AI, as a human design process, is at the pinnacle of the sciences of the artificial. The field of AI is composed of both natural and social sciences in a way that makes the two inextricably tied together. Therefore, solving the challenges of AI through a purely natural science or a purely social science approach will miss the emergent behaviors of the system that requires both working in tandem.

While the differences between the natural and social sciences may look insurmountable, they are, however, complementary in several important ways when brought together through a human-centered design process. Human design is a center piece in the sciences of the artificial and should be at the forefront for how we can approach IS with an emphasis toward the socio-technical.

The natural and social sciences both have a deep appreciation for the scientific method by continuous experimentation and reevaluation of knowledge claims. For instance, these sciences use inductive and deductive arguments to make specific knowledge claims. These scientific disciplines value rigorous scrutiny of their peers and feedback from communities of colleagues.

The similarities and differences between the natural and social sciences also make them complementary under IS for understanding AI. First, because our planet's resources are finite, the natural sciences help inform debates on the environmental impacts of AI, which requires the unique perspectives and empirical rigor of natural scientists to inform debate and policies that impact humans and society. Conversely, social sciences can complement the natural sciences by informing them on the potential unforeseen impacts of supporting the entire AI ecosystem, while also informing the ethical debates that can arise from such discussions.

Second, IS can help these sciences reach consilience by harmonizing the different questions and approaches that arise from debating the merits of AI technologies. For instance, the natural sciences inform the domain of "how" questions about naturalistic phenomenon through empirical research and hypothesis testing. The social sciences provide answers to the "why" questions that the natural sciences cannot answer satisfactorily. Yet, many of the most difficult problems in AI require answers to both these types of questions from these scientific communities to inform a truly interdisciplinary, socio-technical approach within IS.

This organization to IS provides more of a holistic systems approach to problems. The systems approach embodies a problem-solving philosophy that presupposes phenomena result from system behaviors. Therefore, to address any complex phenomena, one must consider not only its internal components, but the environment as well (Mattessich, 1982). By considering sciences of the artificial as part of a larger systems approach brings in the socio-technical perspective necessary to address the many challenges of AI in the field of IS.

## Discussion

IS today is more important than ever for understanding the implications of AI. Humans are ensconcing themselves in more layers of AI technologies than ever before (e.g., Metaverse, Oculus Rift, Apple Vision Pro). While AI encompasses a host of different technologies, it will have far-reaching ramifications for humans and society at large. The need for IS researchers who can bridge the gap between the natural and social sciences will be crucial for helping to inform technology adoption, fair policies, and just laws. IS, though ambiguous in some respects due to its broad interdisciplinary approach to science, can deftly inform research approaches by having a foot planted in both worlds. The philosophical underpinnings of both approaches, whether it is a grounded theory approach by social science researcher or the mathematical rigor of the natural sciences, provide a much richer and comprehensive understanding for informing all aspects of IS and our sociotechnical futures.

Varieties of methods for generating and justifying knowledge claims and the conceptual tools for fully understanding the implications of AI and similar technologies will lead to new insights. Historical neglect of a true socio-technical framing of IS research has likely led to not only the decline of the field, but an impoverished view of the phenomena of interest. Therefore, these reasons compel us to propose an amended definition of IS:

*Information Science is an interdisciplinary design science that recognizes the socio-technical nature of information by applying a systems approach to examine: the generation of information;*

*the behaviors that govern the flow, processing, storage, sharing, and use of information; and how*
*information and its associated technologies impact, human cognition, ecologies, and society.*

The proposed definition improves upon previous definitions by adding additional components of cognition, social, and behavioral aspects that have become increasingly important within IS broadly, and intelligent technologies specifically. These modifications to previous definitions help towards capturing the tensions and paradoxes that humans and societies experience daily when interacting with technologies in various settings. This amended definition provides the additional perspectives researchers need to fully understand IS for the new millennium.

**Conclusion**

The IS field stands at a crossroads today and the choices the field makes for what it pursues may determine the strength of its contribution as a truly interdisciplinary science. Often, fields that call themselves interdisciplinary convey a sense of weakness and may be seen as inferior to the more well-established disciplines (e.g., physics, chemistry, mathematics, engineering) (Buckland, 2012). Worse still, IS has also been maligned as a field of study with claims that it is neither about information nor a true science (Furner, 2015).

Reinvigorating the field of IS will require impactful research that profoundly moves the needle towards better scientific explanations for phenomena of interest. The interdisciplinary nature of IS has prevented it from having what Hjørland (2014) calls a "sufficiently strong centripetal force that can keep the field together" (p. 1). However, recrafting the definition of IS with an emphasis on a socio-technical approach can act as the unifying force for the field. The socio-technical understanding of organizational entities is crucial for understanding how people, technology, and environment interact with one another (Cummings & Markus, 1979). The rapid proliferation of AI and its numerous pitfalls will summarily challenge any narrow interpretation of the phenomenon, which could lead to shortsighted policy and law for curbing its potential harms.

The challenging aspects of AI and the harms it has already wrought on the environment, societies, and individual humans must be addressed to avoid further damages. IS must balance both sides of the technology and human sides of the equation or risk evolving into two separate information sciences (Saracevic, 1999). In the end, amending the definition of IS is a good start, yet it is clearly not enough and far from over.

**Disclosure statement**

The authors report there are no competing interests to declare. The views

**Disclaimer**

The views expressed are those of the authors and do not reflect the official policy or position of the Department of the Navy, Department of Defense or the US Government.